\documentstyle[12pt]{article}
\newcommand{\beq}{\begin{equation}}
\newcommand{\eeq}{\end{equation}}
\newcommand{\bea}{\begin{eqnarray}}
\newcommand{\eea}{\end{eqnarray}}

\begin{document}
\setcounter{page}{0}
\topmargin 0pt
\oddsidemargin 5mm
\renewcommand{\thefootnote}{\fnsymbol{footnote}}
\newpage
\setcounter{page}{0}
\begin{titlepage}
\begin{flushright}
QMW 93-28
\end{flushright}
\begin{flushright}
hep-th/yymmxxx
\end{flushright}
\vspace{0.5cm}
\begin{center}
{\large {\bf The WZNW Model As An Integrable Perturbation Of The
Witten Conformal Point}} \\
\vspace{1.8cm}
\vspace{0.5cm}
{\large Oleg A. Soloviev
\footnote{e-mail: soloviev@V1.PH.QMW.ac.uk}\footnote{Work supported by
S.E.R.C.}}\\
\vspace{0.5cm}
{\em Physics Department, Queen Mary and Westfield College, \\
Mile End Road, London E1 4NS, United Kingdom}\\
\vspace{0.5cm}
\renewcommand{\thefootnote}{\arabic{footnote}}
\setcounter{footnote}{0}
\begin{abstract}
{We show that the WZNW model with arbitrary $\sigma$-model coupling constant
may be viewed as a $\sigma$-model perturbation of the WZNW theory around the
Witten conformal
point. In order for the $\sigma$-model perturbation to be relevant, the level
$k$ of the
underlying affine algebra has to be negative. We prove that in the large $|k|$
limit the
perturbed WZNW system with negative $k$ flows to the conformal WZNW model with
positive level. The flow appears to be integrable due to the existence of
conserved currents satisfying the Lax equation. This fact is in a favorable
agreement with the integrability of the WZNW model discovered by Polyakov and
Wiegmann within the Bethe ansatz technique.}
\end{abstract}
\vspace{0.5cm}
\centerline{October 1993}
 \end{center}
\end{titlepage}
\newpage
%******************************************************************
\section{Introduction}
The Wess-Zumino-Novikov-Witten (WZNW) model [1-4] is a nonlinear two
dimensional $\sigma$-model extended by the Wess-Zumino term [5]. The action of
the theory is written as follows [2]
\begin{equation}
S={1\over4\lambda}\int d^2x \,Tr (\partial_\mu
g\partial^\mu g^{-1})\;+\;k\Gamma,\end{equation}
where $g$ is the matrix field taking its values on the Lie group $G$; $\lambda$
is a $\sigma$-model coupling constant; $\Gamma$ is the Wess-Zumino term [5];
$k$ is a Wess-Zumino coupling constant. For compact groups the Wess-Zumino term
is well defined only modulo $2\pi$ [2, 6], therefore, the parameter $k$ must be
an integer in order for the quantum theory to be single valued with the
multivalued classical action. For noncompact groups the level may be arbitrary.

Polyakov and Wiegmann [6] have solved the WZNW model in eq. (1.1) by means of
the Bethe ansatz technique. However the computation of
correlation functions
remains beyond the powers of the Bethe ansatz method. Much more detailed
information about the WZNW model is obtained at its conformal point [2] within
the current-current algebraic approach which gives rise to the Ward identities
enabling us to compute correlation functions [3]. In spite of the mighty power
of the Knizhnik-Zamolodchikov formalism it can say nothing about the theory
away
from the critical point. Therefore, there are still some reasons for
investigating
the WZNW model with arbitrary $\lambda$.

The aim of the present paper is to present the WZNW model with arbitrary
$\lambda$ as a certain integrable perturbation of the WZNW model around its
conformal point.
We will show that in the theory there is a small parameter which may serve to
tune the coupling $\lambda$ over a large range of values. We find rather
surprisingly that the considered perturbation can be performed only around the
Witten conformal point with negative level $k$. Then the renormalization group
flow takes the WZNW model to the second critical point which is identified
with the
Witten conformal point with positive level $|k|$.
In the framework of the perturbative formulation of the WZNW model
the integrability of the latter uncovered by
Polyakov and Wiegmann within the Bethe ansatz approach originates naturally
from the integrability of the model at the Witten conformal
point possessing an infinite number of conserved currents. It turns out that
the currents of the underlying affine algebra of the
conformal model
beyond the criticality continue to be conserved currents along the flow.
Due to the integrability of the
perturbed theory, the latter can be solved exactly in the framework of the
S-matrix approach [7]. On the other hand, the hope is that the explicit form of
the perturbation might help one to understand correlation functions of the WZNW
model beyond the conformal point.

In section 2 we review the basic features of the WZNW model at the Witten
conformal point. In section 3 we describe the $\sigma$-model perturbation of
Witten's fixed point with negative level $k$. It is proved that the suggested
perturbation obeys all
conditions for relevant and renormalizable perturbations. In section 4 the
coefficient in the fusion rule of the perturbation operator is calculated in
the large $|k|$ limit. In section 5 we discuss the renormalization group flow
between the ultraviolet (starting) and infrared (perturbative) fixed points of
the WZNW model. In section 6 we summaries our results and comments on them

The present paper is an extended version of Ref. 8.

\section{Witten's conformal point}

The WZNW model has achieved much attention during the last decade due to its
special properties at the conformal point discovered by Witten [2]. It has
been realized that the WZNW model at the coupling constant
\begin{equation}
\lambda=\pm|4\pi/k|\end{equation}
can be quantized nonperturbatively either within the algebraic current-current
Hamiltonian approach [3] or by means of the free field representation method in
the path integral formalism [9]. Yet, the WZNW model can be explored also as a
nonlinear sigma model which is conformal at the fixed point [2, 10, 11].

In the present paper we will actively use the properties of the WZNW model at
the Witten conformal point given by eq. (2.2). Therefore, we would like to
sketch some of them in this section.

For our purposes it is more convenient to use complex coordinates
\begin{equation}
z=x^1+ix^2,\;\;\;\;\;\;\;\bar z=x^1-ix^2,\end{equation}
with the metric having the form
\begin{equation}
ds^2=dzd\bar z.
\end{equation}
In these coordinates the action $S^{*}$ of the WZNW model at the Witten
conformal point is written as follows
\begin{equation}
S^{*}(g)=-(k/4\pi)\{\int
d^2z\left[Tr|g^{-1}dg|^2\;+\;(i/3)d^{-1}Tr(g^{-1}dg)^3\right]\}.\end{equation}

The functional $S^{*}(g)$ obeys the Polyakov-Wiegmann formula [6]
\begin{equation}
S^{*}(gh)=S^{*}(g)\;+\;S^{*}(h)\;-\;{k\over2\pi}\int d^2z
Tr(g^{-1}\partial_zg\cdot\bar\partial_{\bar z}hh^{-1}).\end{equation}
By using this formula one can easily establish the symmetry of the theory
$S^{*}$ under the transformations
\begin{equation}
g(z,\bar z)\rightarrow\bar\Omega(\bar z)g(z,\bar z)\Omega,\end{equation}
where $\Omega(z)$ and $\bar\Omega(\bar z)$ are arbitrary $G$-valued matrices
analytically depending on the complex coordinates (2.3).

The symmetry gives rise to an infinite number of conserved currents which can
be derived from the basic currents $J$ and $\bar J$,
\begin{eqnarray}
J&=&J^at^a\,=\,-{1\over 2}kg^{-1}\partial g,\nonumber\\ & & \\
\bar J&=&\bar J^at^a\,=\,-{1\over 2}k\bar\partial gg^{-1}\nonumber,
\end{eqnarray}
satisfying the equations of motion
\begin{equation}
\bar\partial J=0,\;\;\;\;\;\;\partial\bar J=0.\end{equation}
In eqs. (2.8) $t^a$ are the generators of the Lie algebra ${\cal G}$ associated
with the Lie group $G$,
\begin{equation}
[t^a, t^b]=f^{abc}t^c,
\end{equation}
with $f^{abc}$ the structure constants.

Knizhnik and Zamolodchikov [3] have shown that the existence of an infinite
number of conserved currents forming an affine algebra $\hat{\cal G}$, together
with the Virasoro algebra, leads to the Ward identities for correlation
functions. We will make use of the following one
\begin{equation}
\langle J^a(z)\phi_1(z,\bar z_1)\cdot\cdot\cdot\phi_N(z_N,\bar z_N)\rangle=
{\sum^N_{j=1}}{t^a_j\over z-z_j}\langle\phi_1(z_1,\bar
z_1)\cdot\cdot\cdot\phi_N(z_N,\bar z_N)\rangle.\end{equation}
Here the matrices $t^a_i$ correspond to the left representation of the
affine-Virasoro primary fields $\phi_i(z,\bar z)$. The given Ward identity is a
direct consequence of the following operator product expansion (OPE)
\begin{equation}
J^a(w)\phi_i(z,\bar z)={t^a_i\over w-z}\phi_i(z,\bar z)\;+\;reg.,
\end{equation}
which is determined by the transformation property of the field $\phi_i$ with
respect to the infinitesimal transformations
\begin{eqnarray}
\Omega(z)&=&I\;+\;\omega^a(z)t^a,\nonumber\\ & & \\
\bar\Omega(\bar z)&=&I\;+\;\bar\omega^a(\bar z)t^a.\nonumber
\end{eqnarray}
Note that the current $J$ itself is not an affine primary field since its OPE
with itself is
\begin{equation}
J^a(z)J^b(w)={k\delta^{ab}\over(z-w)^2}\;+\;{f^{abc}\over z-w}J^c(w)\;+\;reg.
\end{equation}
The last equation fixes our normalization of the affine currents.

Any affine-Virasoro primary field $\phi_i$ in the WZNW model is degenerate and
its dimensions are given by [3]
\begin{equation}
\Delta_i={c_i\over c_V+k},\;\;\;\;\;\;\bar\Delta_i={\bar c_i\over c_V+k},
\end{equation}
where $c_i=t^a_it^a_i,\;\bar c_i=\bar t^a_i\bar t^a_i$ and $c_V$ is defined
according to
\begin{equation}
f^{acd}f^{bcd}=c_V\delta^{ab}.\end{equation}

\section{The $\sigma$-model perturbation}

It is natural to try to consider the WZNW model around the Witten conformal
point. To this end we present the $\sigma$-model coupling constant in the
following form
\begin{equation}
{1\over\lambda}={1\over\lambda^{*}}\;+\;{\epsilon k^2\over4},\end{equation}
where $\lambda^{*}$ is the Witten conformal point; $\epsilon$ is a small
parameter. With the given reparametrization of $\lambda$ the action of the WZNW
model is written as follows
\begin{equation}
S=S^{*}\;+\;\epsilon\int d^2z\,Tr({k^2\over 4}\partial g\cdot\bar\partial
g^{-1}).\end{equation}
Here $S^{*}$ is the action of the exact conformal theory. So, the second term
in eq. (3.18) appears to be a certain perturbation to the conformal system
$S^{*}$.

At the quantum level we have to define a quantum perturbation associated with
the classical expression in eq. (3.18). Obviously the expansion of the
partition function in the
perturbation around Witten's conformal point allows us to use normal ordering
with respect to the affine currents [12, 13]. To set the perturbation to the
normal ordered form, we use the following identity
\begin{equation}
Tr({k^2\over 4}\partial g\cdot\bar\partial g^{-1})=\phi^{a\bar a}\cdot
J^a\cdot\bar J^{\bar a},\end{equation}
where the currents $J^a,\;\bar J^{\bar a}$ are given by eqs. (2.8) and the
field $\phi^{a\bar a}$ is as follows
\begin{equation}
\phi^{a\bar a}=Tr(g^{-1}t^agt^{\bar a}).\end{equation}
In the theory $S^{*}$ the field $\phi^{a\bar a}$ is both an affine and Virasoro
primary field with scaling dimensions
\begin{equation}
\Delta_\phi=\bar\Delta_\phi={c_V\over c_V+k},\end{equation}
where $c_V$ is as in eq. (2.16).

Thus the normal ordered perturbation can be defined according to the rule
\begin{equation}
O(z,\bar z)\equiv:\phi^{a\bar a}\cdot J^a\cdot \bar J^{\bar a}:(z,\bar z)=
\oint{dw\over2\pi i}\oint{d\bar w\over2\pi i}{J^a(w)\cdot\bar J^{\bar a}(\bar
w)\cdot\phi^{a\bar a}(z,\bar z)\over|z-w|^2}.
\end{equation}
Here in the numerator the product is understood as an OPE. Keeping in mind eq.
(2.12), it is easy to see that the given product does not contain singular
terms.
Correspondingly the WZNW model can be described as a perturbed
conformal model with the action
\begin{equation}
S=S^{*}\;-\;\epsilon\int d^2z \,O(z,\bar z).\end{equation}
Note that the proposed perturbation $O$ is not a product of analytical
operators
{}.

Let us now turn to the large $|k|$ limit. It is not hard to see that in this
limit the perturbation $O$ becomes a quasimarginal operator with anomalous
dimensions
\begin{equation}
\Delta=\bar\Delta=1\;+\;{c_V\over k}\;+\;{\cal O}(k^{-2}).\end{equation}
Hence, when $k$ is negative the given perturbation is to be classified as
relevant; whereas for positive $k$ one should refer to an irrelevant
perturbation. Note that $k$ must be lesser than $-c_V$.

In the case of the relevant perturbation the perturbing operator $O$ possesses
an important property. Namely the OPE of $O$ with itself does not lead to new
relevant operators but $O$. Indeed, the perturbation by the operator $O$ in eq.
(3.23) preserves explicitly the global $G\times G$ symmetry of the affine
$\hat{\cal G}\times\hat{\cal G}$ invariance of the conformal theory $S^{*}$.
Due to the $G\times G$ symmetry, the operator $O$ has to obey the following
fusion rule
\begin{equation}
O\cdot O=[O],\end{equation}
where the square brackets denote the contributions of $O$ and the corresponding
descendants of $O$.

The fusion rule given by eq. (3.25) guarantees the
renormalizability of the perturbed conformal model. Therefore, given the
perturbation one can try to calculate the renormalization beta function
associated with the coupling $\epsilon$.

Away of criticality, where $\epsilon\ne0$, the beta function is defined
according to [14--17]
\begin{equation}
\beta=[2-(\Delta+\bar\Delta]\epsilon\;-\;\pi C\cdot\epsilon^2\;+\;{\cal
O}(\epsilon^3),\end{equation}
where $(\Delta,\;\bar\Delta)$ are given by eq. (3.24). The constant $C$ is
taken here to be the coefficient of the three point function
\begin{equation}
\langle O(z_1)O(z_2)O(z_3)\rangle={C\over|z_{12}|^{\Delta+\bar\Delta}
|z_{13}|^{\Delta+\bar\Delta}|z_{23}|^{\Delta+\bar\Delta}}\end{equation}
when the two point functions are normalized to unity.

One can easily solve equation (3.26) to find fixed points of the beta function.
There are two solutions
\begin{equation}
\epsilon_1=0,\;\;\;\;\;\;\epsilon_2=-2c_V/(\pi Ck).\end{equation}
The first one is nothing but Witten's conformal point of the $S^{*}$ model;
whereas the second solution signifies a new nontrivial conformal point in the
WZNW model. In order to derive the value of the second conformal point, one has
to compute the coefficient $C$ in eq. (3.27).  This task will be addressed in
the next section.

\section{The coefficient C}

Bearing in mind the definition of the operator $O$ (3.22) we obtain the
following expression for the three point function
\begin{eqnarray}
&\langle O(z_1,\bar z_1)O(z_2,\bar z_2)O(z_3,\bar z_3)\rangle=
\oint{dw_1\over2\pi i}\oint{d\bar w_1\over2\pi i}
\oint{dw_2\over2\pi i}\oint{d\bar w_2\over2\pi i}
\oint{dw_3\over2\pi i}\oint{d\bar w_3\over2\pi i}&\nonumber\\ & & \\
&\langle J^a(w_1)\bar J^{\bar a}(\bar w_1)\phi^{a\bar a}(z_1,\bar z_1)
J^b(w_2)\bar J^{\bar b}(\bar w_2)\phi^{b\bar b}(z_2,\bar z_2)
J^c(w_3)\bar J^{\bar c}(\bar w_3)\phi^{c\bar c}(z_3,\bar z_3)
\rangle.&\nonumber\end{eqnarray}
The correlator in the r.h.s. of eq. (3.29) can be readily calculated. By using
the Ward identity given by eq. (2.11) as well as the OPE in eq. (2.14) we find
\begin{equation}
\langle O(z_1,\bar z_1)O(z_2,\bar z_2)O(z_3,\bar z_3)\rangle={k^2f^{abc}f^{\bar
a\bar b\bar c}C_{\phi\phi\phi}^{a\bar a\;b\bar b\;c\bar
c}\over|z_{12}|^{2\Delta}|z_{13}|^{2\Delta}|z_{23}|^{2\Delta}},\end{equation}
where $C_{\phi\phi\phi}$ is the coefficient of the three point function
\begin{equation}
\langle\phi^{a\bar a}(z_1,\bar z_1)\phi^{b\bar b}(z_2,\bar z_2)\phi^{c\bar
c}(z_3,\bar z_3)\rangle={C_{\phi\phi\phi}^{a\bar a\;b\bar b\;c\bar c}\over
|z_{12}|^{2\Delta_\phi}|z_{13}|^{2\Delta_\phi}|z_{23}|^{2\Delta_\phi}}.
\end{equation}

For our aims it is sufficient to estimate $C$ in leading order in $1/k$.
Therefore, we can consider the limit $|k|\rightarrow\infty$. In this limit the
operator
\begin{equation}
K^a=:\phi^{a\bar a}\bar J^{\bar
a}:=Tr(-{k\over2}g^{-1}\bar\partial gt^a) \end{equation}
acquires the canonical dimension (0,1). Hence, in the large $|k|$ limit $K^a$
must behave as a current. In particular, the OPE of
$K^a$ with $\phi^{a\bar a}$ has to be as follows
\begin{equation}
K^a(w,\bar w)\phi^{b\bar b}(z,\bar z)=-{f^{abc}\over\bar w-\bar z}\phi^{c\bar
b}(z,\bar z)\;+\;{\cal O}(1/k).\end{equation}
Equation (4.33) gives rise to the useful relation
\begin{equation}
C_{\phi\phi\phi}^{a\bar a\;b\bar c\;c\bar d}f^{\bar b\bar c\bar
d}=f^{abc}\delta^{\bar a\bar b}\;+\;{\cal O}(1/k).\end{equation}
Taking this identity in eq. (3.30) we obtain the following expression
\begin{equation}
\bar C=k^2f^{abc}f^{\bar
a\bar b\bar c}C_{\phi\phi\phi}^{a\bar a\;b\bar b\;c\bar c}=c_V(k\dim G)^2.
\end{equation}
The relation between $\bar C$ and $C$ is as follows
\begin{equation}
\bar C=||O||^2C,\end{equation}
where $||O||$ is a norm of the operator $O$
\begin{equation}
||O||^2=\langle O(1)O(0)\rangle.\end{equation}
Taking into account the definition of $O$ given by eq. (3.22) we find
\begin{equation}
||O||^2=(k\dim G)^2.\end{equation}
Note that $O$ has a positive norm.

Thus,
\begin{equation}
C=\bar C/(k\dim G)^2=c_V\;+\;{\cal O}(1/k).\end{equation}

\section{A flow}

With the given expression for $C$ the second solution in (3.28) is found as
follows
\begin{equation}
\epsilon_2=-2/(\pi k).\end{equation}
Correspondingly the $\sigma$-model coupling constant $\lambda$ at the second
critical point is given by
\begin{equation}
1/\lambda=(k/4\pi)\;-\;(k/2\pi)=-(k/4\pi).\end{equation}
Thus, we established that the critical WZNW model $S^{*}(k)$ with negative
$k=-|k|$, being perturbed by the $\sigma$-model term, arrives at the critical
WZNW model $S^{*}(l)$ with positive level $l=-k=|k|$. In the case of the
compact group $G$ the critical point $\epsilon_1$ in (3.28) corresponds to the
nonunitary WZNW theory $S^{*}$(k); whereas the second solution $\epsilon_2$
appears to be the unitary system $S^{*}(l)$. Therefore, the flow from
$\epsilon_1$ to $\epsilon_2$ ought to be a unitary flow. This conjecture can be
justified by the Zamolodchikov's c-theorem [18]. According to the theorem, if
the flow between ultraviolet and infrared fixed points is unitary, then the
Virasoro central charge corresponding to the perturbative (infrared) fixed
point must be lesser than the central charge at the ultraviolet conformal
point. In the case under consideration, we know $c(\epsilon_1),\;c(\epsilon_2)$
exactly. It is easy to check that
\begin{equation}
c(\epsilon_2)\;-\;c(\epsilon_1)={-2kc_V\dim G\over c_V^2-k^2}.
\end{equation}
Now it becomes clear that when $k<-c_V$ the difference written above is less
than zero in full agreement with the c-theorem. Interestingly, in the limit
$k\rightarrow-\infty$ the critical points $\epsilon_1,\;\epsilon_2$ collide so
that the corresponding central charges become equal.

The point to be made is that the Virasoro central charge at the perturbative
conformal point $\epsilon_2$ might be also estimated by
perturbation theory. There is available, for example, the Cardy-Ludwig formula
[14-16]. However, in the case under consideration there are two obstacles
obstructing the use of this formula. The first one is that the perturbation
operator $O$ given by eq. (3.22) is not a primary operator; it belongs to
$[\phi^{a\bar a}]$. Namely,
\begin{equation}
O=J^a_{-1}\bar J^{\bar a}_{-1}\phi^{a\bar a}.
\end{equation}
The second problem is that the c-function may turn out to be a nonanalytical
function of $\epsilon$ in the interval $]\epsilon_1,\;\epsilon_2[$. Indeed, in
this interval the coupling constant $1/\lambda$ changes its sign
and, hence, passes through zero so that $\lambda\rightarrow\pm\infty$.
Therefore, the perturbative formula for the Virasoro central charge needs
to be carefully investigated.

It is instructive to compute anomalous dimensions of the operator $O$ at the
second
conformal point. We can use the perturbative formula due to Redlich [19]. To
given order in $1/k$ the formula yields
\begin{equation}
\Delta(\epsilon_2)=\bar\Delta(\epsilon_2)=1\;-\;c_V/k.\end{equation}
The given expression is consistent with the exact result
\begin{equation}
\Delta(\epsilon_2)=\bar\Delta(\epsilon_2)=1\;+\;{c_V\over c_V-k},
\end{equation}
where $k$ is negative.

As the level $k$ goes to $-\infty$, the $\sigma$-model coupling constant
$\lambda$ is very sensitive to small changes
of the perturbative parameter $\epsilon$. Therefore, by doing the
perturbation with $\epsilon$ being smaller than the critical value
$\epsilon_2$, we should be able to gain insight into the behaviour of the WZNW
model with arbitrary $\lambda$ in the intervals $]-\infty,\;-4\pi/|k|[$ and
$]4\pi/|k|,\;+\infty[$.

Let us look at the integrability of the WZNW model in the intervals just
mentioned above. Once we are dealing with the perturbation around conformal
model, we can try to disclose the existence of nontrivial conserved currents in
the model by following in the manner of Ref. [20]. Namely, we may try to
discover
what is happening to the conserved currents of the unperturbed model in the
course
of the perturbation. In the WZNW model at the critical point $\epsilon_1$
there are
infinite number of conserved currents forming the affine algebra. Away of the
criticality the affine currents are no longer analytical but yet may continue
to be
conserved. To clarify this point, we have to consider the following correlator
\begin{equation}
\langle J^a(z,\bar z).\,.\,.\rangle=\langle J^a(z,\bar z).\,.\,.
\rangle_{S^{*}}\;+\;
\epsilon\int\langle J^a(z)O(z_1,\bar z_1).\,.\,.\rangle_{S^{*}}d^2z_1\;+\;{\cal
O}(\epsilon^2).\end{equation}
When $(z_1,\bar z_1)$ approaches the point $(z,\bar z)$ the integral in eq.
(5.46) should be regulated. When this is done , $\partial_{\bar z}J^a$ is no
longer zero. By using the definition of the operator $O$ given by eq. (3.22)
we obtain the formula
\begin{equation}
\int\langle J^a(z)O(z_1,\bar z_1).\,.\,.\rangle_{S^{*}}d^2z_1=\epsilon
k\;\int\langle{K^a(z_1,\bar z_1)\over(z-z_1)^2}.\,.\,.\rangle_{S^{*}}d^2z_1,
\end{equation}
where $K^a$ is as in eq. (4.32). The last equation gives rise to the following
relation
\begin{equation}
\bar\partial J^a=-\eta\;\partial K^a\;+\;{\cal O}(\epsilon^2),
\end{equation}
where
\begin{equation}
\eta=\pi\epsilon k/2           \end{equation}
So, to first order in $\epsilon$ the current $J$ continues to be conserved.
Moreover, it satisfies the Lax equation
\begin{equation}
\bar\partial J=\pi\epsilon\;[K,J].
\end{equation}
This establishes the integrability of the WZNW model to lowest order in the
perturbation.

When $\epsilon$ approaches the critical value $\epsilon_2$, one has to go to
higher orders in $\epsilon$ since the ratio $\epsilon k$ is no longer small but
goes to a finite constant. Nevertheless, we can argue that the current $J$
must be conserved up to the fixed point $\epsilon_2$. Indeed, Polyakov and
Wiegmann [6] have managed to prove the integrability of the WZNW model with
arbitrary $\lambda$ within the Bethe ansatz technique. Therefore, the system
has to possess an infinite number of conserved currents. At the second
critical point the model becomes again a conformal system with an affine
symmetry. Of course, it is very interesting to realize the deformation of the
affine algebra under the perturbation.

\section{Conclusions}

We have seen that the WZNW model with arbitrary $\sigma$-model coupling
constant $\lambda$ can be presented as the conformal WZNW model perturbed by
the $\sigma$-model term. The perturbation turns out to be relevant around the
Witten
conformal point provided the level underlying affine algebra is negative. In
the
course of the perturbation, the model arrives at the infrared fixed point
corresponding to the conformal WZNW model with positive level. The flow from
the nonunitary conformal point to the unitary conformal point appears to be
integrable. It is rather amusing that the smaller the size of the perturbation
one takes, the more information of the WZNW model one can get. In the limit
$|k|$ goes to infinity the theory seems to be solvable for the whole line of
values of the
coupling constant $\lambda$.

Given the perturbative description of the WZNW model one can try to explore
correlation functions beyond the Witten conformal point. In this connection it
is very important to understand the transition from the
nonunitary phase to the unitary phase.

We expect also that within the perturbative formulation it will be possible to
study the deformation of the affine algebra along the integrable
renormalization group flow.

\par \noindent
{\em Acknowledgement}: I would like to thank J. M. Figueroa-O'Farrill,
M. Green, C. Hull, G. Mussardo, E. Ramos, W. Sabra, S. Thomas and I. Vaysburd
for useful
discussions. I would also like to thank the SERC for financial support.


\begin{thebibliography}{99}
\bibitem{1} S. P. Novikov, Usp. Mat. Nauk {\bf 37} (1982) 3.
\bibitem{2} E. Witten, Commun. Math. Phys. {\bf 92} (1984) 455.
\bibitem{3} V. G. Knizhnik and A. B. Zamolodchikov, Nucl. Phys. {\bf B247}
(1984) 83.
\bibitem{4} D. Gepner and E. Witten, Nucl. Phys. {\bf B278} (1986) 493.
\bibitem{5} J. Wess and B. Zumino, Phys. Lett. {\bf 37B} (1971) 95.
\bibitem{6} A. M. Polyakov and P. B. Wiegmann, Phys. Lett. {\bf B131} (1983)
121; {\bf 141} (1984) 223; P. B. Wiegmann, JETP Lett. {\bf 39} (1989) 180.
\bibitem{7} A. B. Zamolodchikov and Al. B. Zamolodchikov, Ann. of Phys. {\bf
120} (1979) 253.
\bibitem{8} O. A. Soloviev, {\it The WZNW model by a perturbation of Witten's
conformal solution}, Preprint QMW-93-25, hep-th/9310014.
\bibitem{7} A. Gerasimov, A. Marshakov, A. Morozov, M. Olshanetsky and S.
Shatashvili, Int. J. Mod. Phys. {\bf A5} (1990) 2495.
\bibitem{6} M. Bos, Ann. of Phys. {\bf 181} (1988) 177; B. de Wit, M. T.
Grisaru and P. van Nieuwenhuizen, {\it The WZNW
model at two loops}, THU-93-15, ITP-SB-93-353, BRX-TH-349, hep-th/9307027.
\bibitem{11} A. A. Tseytlin, {\it Conformal sigma models corresponding to
gauged Wess-Zumino-Witten theories}, Preprint CERN-TH. 6804/93, hep-th/9302083.
\bibitem{12} R. Dashen and Y. Frishman, Phys. Rev. {\bf D 11} (1975) 2781.
\bibitem{13} K. Johnson, Nuovo Cimento {\bf 20} (1961) 773.
\bibitem{10} K. J. Wilson and J. Kogut, Phys. Rep. {\bf 12C} (1974) 75; J. M.
Kosterlitz, J. Phys. C. Solid State {\bf 7} (1974) 1046.
\bibitem{11} J. L. Cardy and A. W. W. Ludwig, Nucl. Phys. {\bf B285} (1987)
687.
\bibitem{12} J. L. Cardy, {\it in} Phase transition and critical phenomena, v.
11, eds. C. Domb and J. L. Lebowitz (Academic Press, 1987).
\bibitem{17} A. B. Zamolodchikov, Sov. J. Nucl. Phys. {\bf 44} (1986) 529.
\bibitem{9} A.B. Zamolodchikov, JETP Lett. {\bf 43} (1986) 730.
\bibitem{13} A. N. Redlich, Phys. Lett. {\bf B217} (1989) 129.
\bibitem{20} A. B. Zamolodchikov, Int. J. Mod. Phys. {\bf A3} (1988) 743.
\end{thebibliography}
\end{document}